\documentclass[onecolumn]{aa}
\usepackage{color}
\usepackage{graphicx}
\begin{document}
\title{The effect of dust obscuration in RR Tel on optical and IR long-term photometry
and Fe II emission lines}

      \author{D. Kotnik-Karuza
   \inst{1}
   \and
   M. Friedjung
   \inst{2}
   \and
   P.A. Whitelock
   \inst{3}
   \and
   F. Marang
   \inst{3}
   \and
   K. Exter
   \inst{4}
   \and
   F.P. Keenan
   \inst{5}
   \and
   D.L. Pollacco
   \inst{5}
          }

   \offprints{D. Kotnik-Karuza}

   \institute{Physics Department, Faculty of Arts and Sciences, Omladinska 14, HR-51000 Rijeka, Croatia\\
              \email{dubravka.karuza-kotnik@ri.htnet.hr}
         \and
             Institut d'Astrophysique de Paris UMR7075 CNRS, Universit\'e Pierre \& Marie Curie, 98 bis Boulevard Arago, 75014 Paris, France\\
             \email{friedjung@iap.fr}
         \and
             South African Astronomical Observatory, Observatory, South Africa and national Astrophysics and Space
Science Programme, University of Cape Town, 7701 Rondebosch, South
Africa
\\
             \email{paw@saao.ac.za}
             \and
             Instituto de Astrofisica de Canarias, La Laguna (Tenerife), Espana\\
             \email{katrina@iac.es}
         \and
             Dept. of pure and applied physics, Queens
       University, Belfast, Northern Ireland \\
       \email{F.Keenan@qub.ac.uk}
             }

      \date{Received date / Accepted date}


  \abstract
   {  }
   {Infrared and optical photometric and spectroscopic observations of the
   symbiotic nova RR Tel are used to study the effects and properties of dust in
   symbiotic binaries containing a cool Mira component, as well as showing "obscuration events"
   of increased absorption, which are typical for such Miras.}
   {A set of photometric observations of the symbiotic nova \object{RR Tel}
in different wavelength bands - visual from 1949 to 2002 and
near-infrared ($JHKL$) from 1975 to 2002 - are presented. The
variability due to the normal Mira pulsation was removed from the
JHKL data, which were then compared with the American Association
of Variable Star Observers' ($AAVSO$) visual light curve. The
changes of the Fe II emission line fluxes during the 1996-2000
obscuration episode were studied in the optical spectra taken with
the Anglo-Australian telescope. }
   {We discuss the three periods during which the Mira component was
heavily obscured by dust as observed in the different wavelength
bands. A change in the correlations of J with other infrared
magnitudes was observed with the colour becoming redder after
JD2446000. Generally, J-K was comparable, while K-L was larger
than typical values for single Miras. A distance estimate of 2.5
kpc, based on the IR data, is given. A larger flux decrease for
the permitted than for the forbidden Fe II lines, during the
obscuration episode studied, has been found. There is no evidence
for other correlations with line properties, in particular with
wavelength, which suggests obscuration due to separate optically
thick clouds in the outer layers.}

{}


\authorrunning{Kotnik-Karuza et al.}
\titlerunning{ Dust obscuration in RR Tel }
   \maketitle
%

\section{Introduction}

   The symbiotic nova RR Tel is an interacting binary, consisting of
a Mira-type cool component, a hot component thought to be a white
dwarf, an extended nebular envelope and a dust envelope around the
cool component. Since its nova-like optical outburst in 1944 long
term spectroscopic and photometric observations have been carried
out in the UV, optical and IR spectral regions in order to
establish a satisfactory model of this puzzling object.
Inter-comparison of data from observations in different wavelength
regions made a significant contribution to these attempts (Heck
and Manfroid \cite{He85}; Contini and Formiggini \cite{Co99};
Penston et al. \cite{Pe83}; Whitelock \cite{Wh87}).

RR Tel belongs to the class of symbiotic Miras, where the cool
giant is a Mira variable (Whitelock \cite{Wh03}). Normal single
Mira variables are pulsating stars at the top of the asymptotic
giant branch (AGB) with periods in the 100 to more than 2000 days
range. They have strong winds, which are thought to be first
levitated by pulsations, and then accelerated by the action of
radiation pressure on the abundant dust, which will then drag the
gas. The dust can also produce ``obscuration events" of increased
absorption, which are rare in single oxygen-rich Miras, but which
occur for the cool components of most symbiotic Miras. Mira
variables can be either oxygen- or carbon-rich, RR Tel being a
member of the oxygen-rich class.

From the evolutionary point of view a low or intermediate mass
star will evolve along the AGB before it becomes a white dwarf.
Symbiotic Miras, unlike other symbiotic binaries, have large
separations and corresponding long orbital periods of probably not
less than 20 years, known orbital periods of other symbiotic
binaries being usually less than 4 years (Belczynski et al.
\cite{Be00}). However, the winds are so strong, that accretion by
the compact component can still be high, in spite of the large
separation. An accreting white dwarf can then undergo continuous
thermonuclear burning of the accreted material or occasional
epochs of such burning. The latter, sporadic thermonuclear events,
are thought to explain the nova like outbursts.

In our present work, we have reanalyzed infrared and optical
photometry, as well as emission line fluxes. As we wish in
particular to understand better the obscuration events, we used
the infrared photometry of RR Tel and compared it with that of
normal isolated Miras (Whitelock et al. \cite{Wh94}, Whitelock et
al. \cite{WM00}, Le Bertre \cite{Be93}, Smith \cite{Sm03}) as well
as with Miras accompanied by hot components in symbiotics
(Whitelock \cite{Wh87}, Whitelock \cite{Wh88}, Whitelock
\cite{Wh03}).
\par

Infrared photometry of RR Tel was published by Feast et al. (
\cite{FW83}), Feast et al. (\cite{Fe77}), Whitelock ( \cite{Wh87})
and Whitelock ( \cite{Wh98}). The J magnitudes up to 2002 were
illustrated in Whitelock (\cite{Wh03}). Here we use the whole
available dataset to construct the JHKL light curves and compare
them with the visual light curve obtained for the same epochs. The
colour changes of RR Tel up to 2002 have also been investigated.

In this work we also show the visual light curve of RR Tel
covering the time interval 1949-2002. The light curve of Contini
and Formiggini ( \cite{Co99}), taken from different sources, ends
in 1995. Recent observations, from 1995 on, are also interesting
because there is evidence of a systematic decrease of the optical
line fluxes from 1996 to 2000.

Such behaviour was observed for the Fe II lines ( Kotnik-Karuza et
al. \cite{KK03}) as well as for the fluxes of other ions measured
in the optical spectra during the period mentioned ( Kotnik-Karuza
et al. \cite{KK04}, Kotnik-Karuza et al. \cite{KF04}). No evidence
of fading was found (Kotnik-Karuza et al. \cite{KK02}) between the
July 1993 observations described by McKenna et al. (\cite{Mc97})
and the July/August 1996 observations described by Crawford et al.
(\cite{Cr99}). We do not have information, which could enable us
to make a comparison in other wavelength regions.

Searching for a possible relationship between the IR fading and the
spectroscopic changes of the hot source, we concentrate
here on the Fe II and [FeII] fluxes, less susceptible than fluxes
of lines of more ionized species formed nearer the hot component,
to be affected by changes of the temperature of the hot component
(M\"urset and Nussbaumer \cite{MN94}).

The distance of RR Tel is not well defined, estimates of it ranging
between about 2.5 kpc (Thackeray \cite{Th77}; Penston et al.
\cite{Pe83}) to 3.6 kpc (Feast et al. \cite{FW83}). By use of more
recent material and with the considerably increased amount of
observational IR data for RR Tel and other symbiotic Miras, we
give a revised distance estimate for RR Tel.

\section{Observations and methods of analysis}

\begin{table*}
\caption{JHKL magnitudes of RR Tel from 1975 to 2002}
\label{table:1}
{\tiny
\begin{tabular}
[c]{lllllllllllllllll}\hline\hline JD-2440000 & J & H & K & L &
JD-2440000 & J & H & K & L & JD-2440000 & J & H & K & L\\\hline
2653.50& 5.84  & 4.94  & 4.21  & 3.17  & 5215.34 & 5.590 & 4.688 & 3.988 &  2.919 &  8067.60 & 6.225 & 4.918 & 3.950 & 2.590\\
2689.50& 6.15  & 5.10  & 4.38  & 3.29  & 5251.27 & 5.329 & 4.319 & 3.662 &  2.667 &  8086.58 & 6.295 & 5.008 & 4.042 & 2.674\\
2881.50& 6.73  & 5.60  & 4.74  & 3.37  & 5481.63 & 6.238 & 5.196 & 4.420 &  3.344 &  8115.49 & 6.526 & 5.230 & 4.254 & 2.907\\
2912.50& 6.35  & 5.34  & 4.48  & 3.29  & 5543.43 & 6.223 & 5.180 & 4.404 &  3.274 &  8142.42 & 6.720 & 5.440 & 4.445 & 3.052\\
2971.50& 5.38  & 4.64  & 4.00  & 3.05  & 5546.45 & 6.173 & 5.133 & 4.365 &  3.259 &  8389.68 & 5.943 & 4.705 & 3.793 & 2.464\\
2995.50& 5.43  & 4.47  & 3.84  & 2.94  & 5565.38 & 6.155 & 5.107 & 4.336 &  3.247 &  8452.45 & 6.060 & 4.757 & 3.852 & 2.588\\
3012.50& 5.40  & 4.42  & 3.82  & 2.90  & 5570.36 & 6.090 & 5.080 & 4.314 &  3.206 &  8500.43 & 6.527 & 5.172 & 4.201 & 2.846\\
3051.50& 5.46  & 4.49  & 3.88  & 3.03  & 5595.34 & 5.845 & 4.898 & 4.179 &  3.059 &  8787.59 & 7.166 & 5.622 & 4.338 & 2.716\\
3062.50& 5.55  & 4.57  & 3.96  & 3.10  & 5614.32 & 5.649 & 4.677 & 3.979 &  2.904 &  8817.53 & 6.939 & 5.443 & 4.262 & 2.723\\
3083.50& 5.66  & 4.65  & 4.08  & 3.18  & 5627.31 & 5.593 & 4.579 & 3.897 &  2.830 &  8854.40 & 6.819 & 5.398 & 4.298 & 2.806\\
3284.68& 6.09  & 5.18  & 4.40  & 3.31  & 5638.29 & 5.443 & 4.448 & 3.803 &  2.818 &  8875.39 & 6.876 & 5.469 & 4.391 & 2.905\\
3322.55& 5.62  & 4.74  & 4.09  & 3.09  & 5849.61 & 6.237 & 5.186 & 4.439 &  3.399 &  8932.30 & 7.255 & 5.921 & 4.828 & 3.257\\
3346.45& 5.46  & 4.53  & 3.88  & 2.99  & 5873.60 & 6.271 & 5.223 & 4.468 &  3.448 &  9146.69 & 6.420 & 5.195 & 4.199 & 2.753\\
3386.38& 5.33  & 4.37  & 3.77  & 2.97  & 5893.52 & 6.210 & 5.189 & 4.444 &  3.442 &  9171.50 & 6.182 & 4.969 & 4.036 & 2.656\\
3412.32& 5.34  & 4.36  & 3.78  & 2.91  & 5906.59 & 6.159 & 5.130 & 4.394 &  3.402 &  9215.47 & 5.922 & 4.731 & 3.866 & 2.599\\
3431.31& 5.45  & 4.45  & 3.85  & 3.05  & 5924.42 & 6.092 & 5.079 & 4.353 &  3.377 &  9586.50 & 5.675 & 4.471 & 3.659 & 2.514\\
3439.32& 5.41  & 4.44  & 3.87  & 3.04  & 5957.35 & 6.144 & 5.131 & 4.391 & 99.999 &  9640.33 & 5.681 & 4.480 & 3.702 & 2.588\\
3644.64& 6.49  & 5.56  & 4.68  & 3.42  & 6029.27 & 5.429 & 4.461 & 3.829 &  2.844 &  9890.70 & 6.125 & 4.959 & 4.022 & 2.675\\
3655.64& 6.41  & 5.48  & 4.60  & 3.38  & 6225.55 & 6.138 & 5.062 & 4.351 &  3.409 &  9937.46 & 5.725 & 4.528 & 3.658 & 2.442\\
3662.67& 6.40  & 5.44  & 4.59  & 3.46  & 6266.49 & 6.255 & 5.191 & 4.448 &  3.489 & 10006.39 & 5.460 & 4.289 & 3.505 & 2.392\\
3675.65& 6.20  & 5.32  & 4.50  & 3.35  & 6351.32 & 5.815 & 4.870 & 4.166 &  3.163 & 10016.27 & 5.481 & 4.313 & 3.542 & 2.433\\
3695.58& 5.93  & 5.03  & 4.27  & 3.17  & 6597.56 & 6.366 & 5.229 & 4.414 &  3.347 & 10212.66 & 6.738 & 5.486 & 4.458 & 3.067\\
3706.55& 5.78  & 4.90  & 4.17  & 3.06  & 6645.50 & 6.928 & 5.740 & 4.789 &  3.522 & 10256.62 & 6.725 & 5.458 & 4.397 & 3.004\\
3729.43& 5.67  & 4.73  & 4.02  & 2.98  & 6693.40 & 6.988 & 5.675 & 4.743 &  3.371 & 10299.50 & 6.482 & 5.102 & 4.004 & 2.629\\
3819.26& 5.42  & 4.41  & 3.83  & 2.89  & 6704.36 & 6.919 & 5.721 & 4.713 &  3.347 & 10590.65 & 7.951 & 6.371 & 4.976 & 3.278\\
3821.26& 5.49  & 4.48  & 3.88  & 2.89  & 6748.32 & 6.718 & 5.576 & 4.600 &  3.242 & 10673.35 & 7.290 & 5.678 & 4.348 & 2.780\\
4070.53& 5.61  & 4.70  & 4.05  & 3.06  & 6923.65 & 6.168 & 4.912 & 4.051 &  2.872 & 10995.51 & 7.864 & 6.498 & 5.173 & 3.392\\
4096.48& 5.27  & 4.37  & 3.77  & 2.85  & 6959.58 & 6.660 & 5.346 & 4.404 &  3.140 & 11057.42 & 7.516 & 6.060 & 4.695 & 2.946\\
4364.62& 6.35  & 5.30  & 4.51  & 3.43  & 6985.51 & 7.064 & 5.726 & 4.727 &  3.359 & 11104.55 & 7.174 & 5.718 & 4.430 & 2.783\\
4429.57& 6.16  & 5.16  & 4.37  & 3.29  & 6999.52 & 7.293 & 5.940 & 4.906 &  3.442 & 11389.45 & 7.694 & 6.322 & 5.106 & 3.395\\
4434.56& 6.16  & 5.16  & 4.36  & 3.26  & 7025.39 & 7.688 & 6.333 & 5.182 &  3.596 & 11417.44 & 7.621 & 6.246 & 5.012 & 3.285\\
4445.48& 6.05  & 5.05  & 4.28  & 3.21  & 7072.37 & 8.093 & 6.657 & 5.307 &  3.573 & 11685.60 & 7.867 & 6.468 & 5.234 & 3.534\\
4744.65& 6.247 & 5.254 & 4.423 & 3.323 & 7331.55 & 7.435 & 5.905 & 4.623 &  3.014 & 11712.62 & 7.919 & 6.518 & 5.267 & 3.569\\
4771.62& 6.232 & 5.222 & 4.357 & 3.228 & 7357.45 & 7.762 & 6.207 & 4.892 &  3.240 & 11743.53 & 7.760 & 6.373 & 5.149 & 3.457\\
4772.53& 6.192 & 5.194 & 4.321 & 3.156 & 7379.50 & 8.001 & 6.491 & 5.126 &  3.394 & 11783.42 & 7.597 & 6.232 & 5.018 & 3.320\\
4777.56& 6.174 & 5.187 & 4.319 & 3.173 & 7393.38 & 8.200 & 6.688 & 5.323 &  3.532 & 11814.38 & 6.898 & 5.646 & 4.542 & 2.944\\
4781.54& 6.178 & 5.204 & 4.333 & 3.196 & 7394.36 & 8.229 & 6.687 & 5.293 &  3.528 & 11828.33 & 6.674 & 5.419 & 4.364 & 2.835\\
4803.54& 6.173 & 5.160 & 4.267 & 3.135 & 7438.25 & 8.582 & 7.041 & 5.554 &  3.634 & 11860.24 & 6.408 & 5.146 & 4.141 & 2.731\\
4815.45& 6.141 & 5.159 & 4.265 & 3.095 & 7657.63 & 7.284 & 5.699 & 4.364 &  2.702 & 12066.63 & 7.783 & 6.409 & 5.257 & 3.614\\
4822.49& 6.071 & 5.073 & 4.201 & 3.096 & 7690.61 & 7.411 & 5.816 & 4.506 &  2.854 & 12087.59 & 7.945 & 6.497 & 5.286 & 3.632\\
4899.31& 5.113 & 4.199 & 3.534 & 2.591 & 7719.55 & 7.537 & 5.978 & 4.681 &  3.006 & 12117.52 & 7.970 & 6.514 & 5.252 & 3.571\\
5100.66& 6.185 & 5.119 & 4.345 & 3.366 & 7744.52 & 7.709 & 6.160 & 4.876 &  3.177 & 12155.44 & 7.868 & 6.423 & 5.107 & 3.358\\
5157.52& 5.933 & 4.932 & 4.181 & 3.148 & 7776.39 & 7.916 & 6.392 & 5.119 &  3.417 & 12183.38 & 7.555 & 6.135 & 4.871 & 3.197\\
5193.43& 5.882 & 4.911 & 4.168 & 3.048 & 7811.35 & 8.004 & 6.490 &
5.259  &  3.558 & 12208.25 & 7.143 & 5.744 & 4.543 & 2.912\\
\hline

\end{tabular}
}
\end{table*}

   \begin{figure}
      \resizebox{\hsize}{!}{\includegraphics[angle=-90]{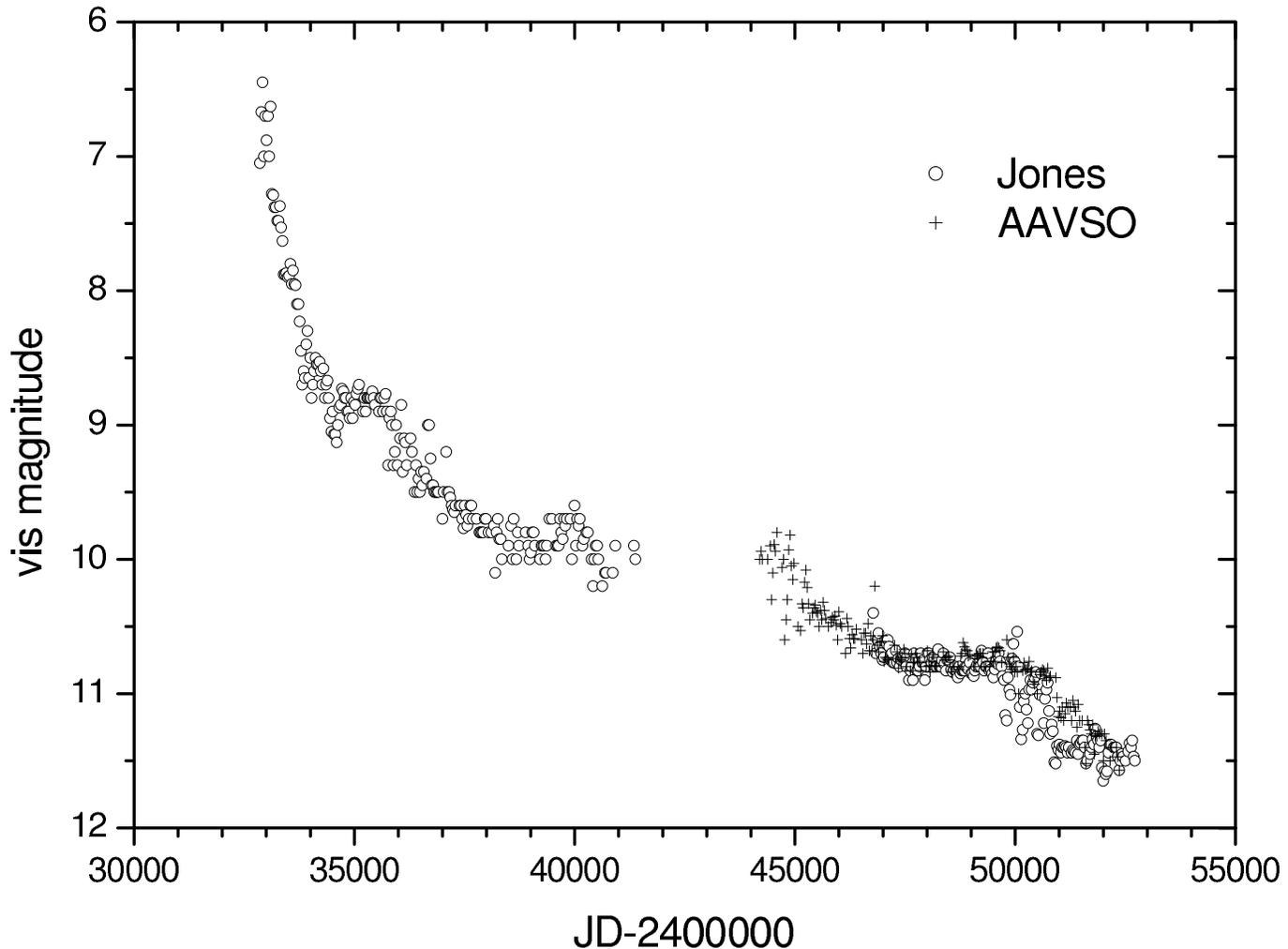}}
      \caption{RR Tel averaged visual estimates over 30 days for the years
1949-2002. Crosses refer to AAVSO observational data, collected
from several observers and given to us by the late J.Mattei. Circles
correspond to the observations of Albert Jones.}
              \label{fig:prv}
    \end{figure}

\begin{figure}
      \resizebox{\hsize}{!}{\includegraphics[angle=-90]{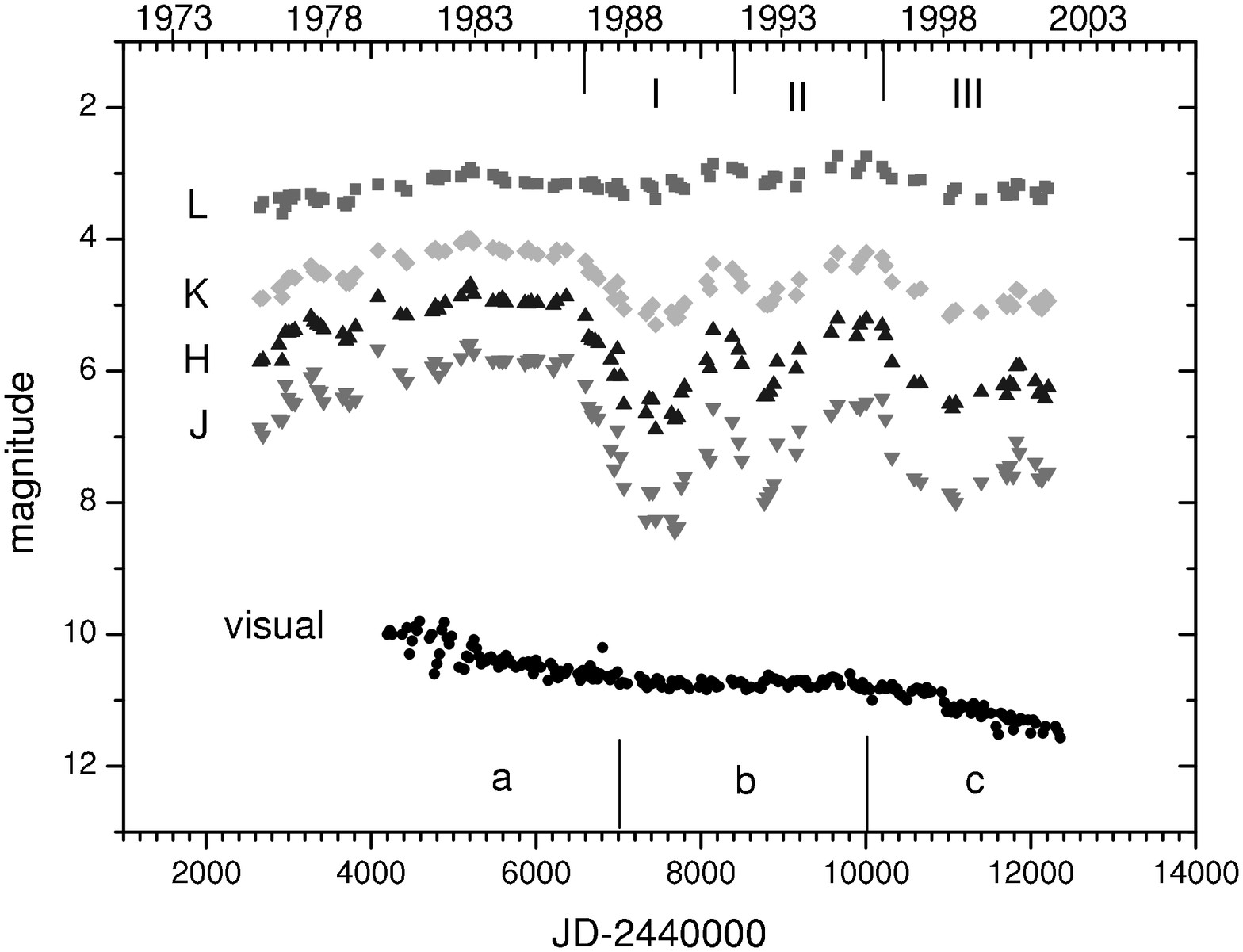}}
      \caption{Binned IR and visual light curves from 1975 to 2002. The
  epochs I, II and III refer to three obscuration events, while a,
  b and c mark different segments of the visual light curve (see
  text)}
              \label{fig:dva}
    \end{figure}

   The visual magnitudes of RR Tel were measured by the AAVSO
observers from the Southern Hemisphere between the years 1949 and
2002. Fig.~\ref{fig:prv} shows the smoothed visual light curve
which was obtained by taking means of magnitudes over 30 day
intervals.

The JHKL magnitudes monitored from the South African Astronomical
Observatory (SAAO) from 1975 to 2002 are given in Table 1.
Infrared JHKL light curves of RR Tel are affected by Mira
pulsations with a period of 387 days (Feast et al. \cite{FW83})
and by long-term variations as observed in many symbiotic Miras
(Whitelock 1987, Mikolajewska et al. \cite{Mi99}; Whitelock
\cite{Wh03}). In order to isolate the long-term trends, we have
corrected the light curves for Mira pulsations by a procedure in
which the observations were binned at the same Mira phase,
taking means of the magnitudes in each IR band over a tenth of the
Mira period. The points were plotted separately at the same phase
within each of these bins. Then we superposed the
curves at different phases by shifting them vertically with
respect to the curve with the best distribution of observations to
obtain a minimum deviation by a least squares fit. The resulting
JHKL curves corrected for Mira pulsations are shown in
Fig.~\ref{fig:dva} together with the overlapping part of the
visual light curve for comparison. Let us note that the JHKL
magnitudes are accurate to better than 0.03 mag in JHK and 0.05
mag at L. Some of the early data, which had been published in Feast
et al. (\cite{FW83}), were subsequently slightly corrected to the
SAAO system as defined by Carter (1990).

In addition, flux calibrated optical spectra, taken with the
Anglo-Australian telescope in 1996 and 2000, are compared. The
former described by Crawford et al. (\cite{Cr99}), covering the
region from 3100-9800 \AA, was obtained on July 22 1996 with a
resolution of about 50000 and was flux calibrated on August 2
1996. The latter was taken in July 2000 with almost twice the
spectral resolution and was flux calibrated with two other
spectra, including one taken with the HST in October 2000.

\section{ Results and discussion}

\bigskip

\subsection{Comparison of visual and IR data}

In all four smoothed IR light curves in Fig.~\ref{fig:dva}, what
we call the events I, II and III, can clearly be resolved as
outstanding features, indicating that during these periods the
Mira was heavily obscured by dust. The amplitudes of the
variations are lower at the longer wavelengths. In addition, there
are clear differences between the L and the other IR light curves,
the L mag being approximately constant with only very slight
obscuration features.

There is no need to correct the AAVSO curve for Mira pulsations
which are not seen in the optical region, because the unobscured
parts of the nebula are the dominant source of emission at these
short wavelengths. Let us recall that the nebulae within the
system have a very complex structure (Nussbaumer \cite{Nu00}). In
the part of the AAVSO curve shown in Fig.~\ref{fig:prv} there are
three separate segments in time intervals a, b and c. The fadings
over intervals a and c are at approximately the same rate, in
contrast to the period b during which the visual magnitude
remained almost constant. In general, the AAVSO curve shows very
little relation to the IR variations. The dust obscuration event
starting about JD 2450000 may be connected with the fading visual
luminosity, the effect being larger in J than in other near-IR
bands.

The changes in the IR and visual light curves are summarized in
the Table 2. The depth of the obscuration in
JHK and L is measured with respect to the pre-obscuration level
during 1983 through to 1986 with the assumption that it reasonably
represents the mean brightness of the Mira. Clearly the assumption
is better for the first obscuration phases than for the later
ones.

\subsection{JHKL colours}

\begin{figure}
      \resizebox{\hsize}{!}{\includegraphics[angle=-90]{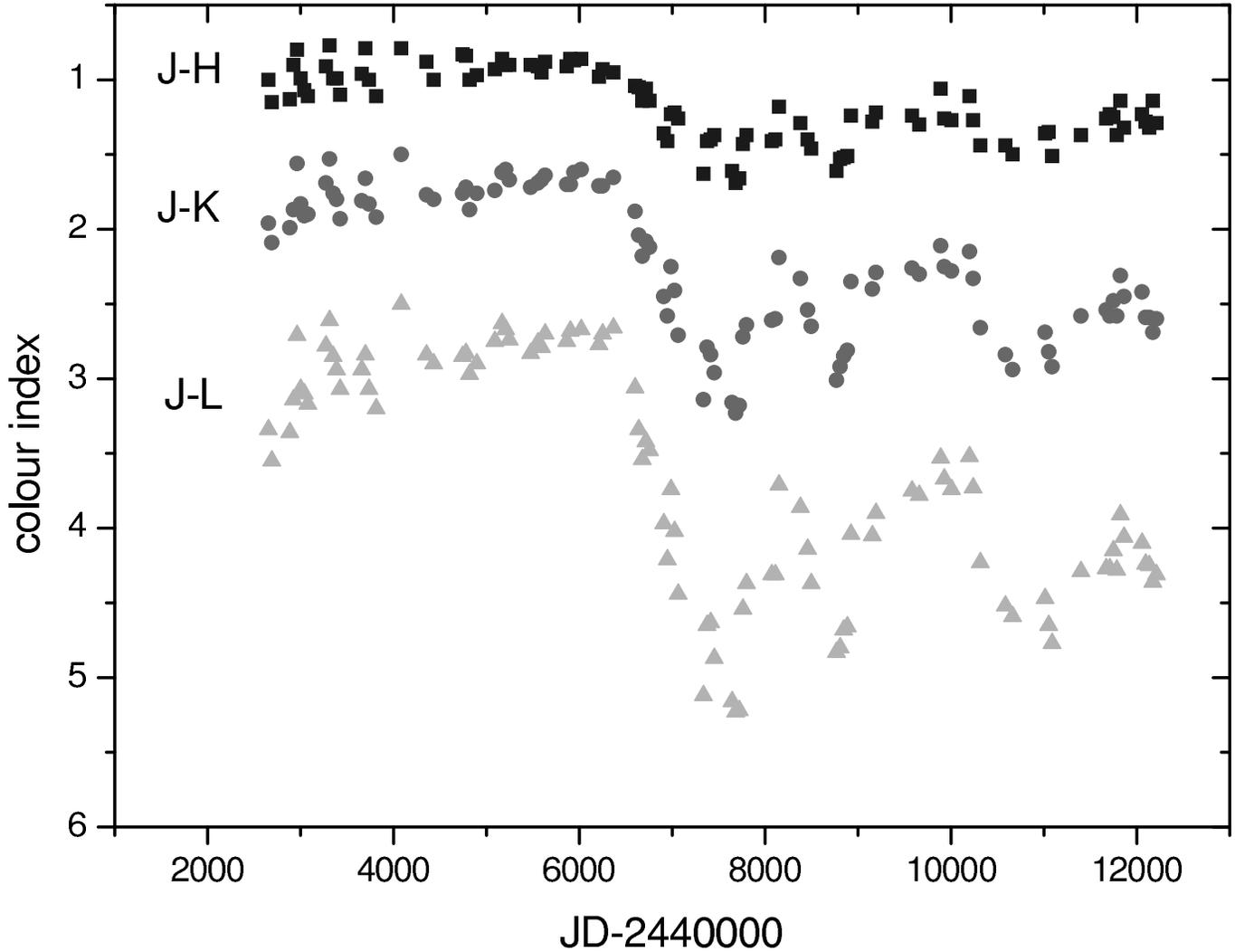}}
      \caption{Time evolution of the J-H, J-K and J-L colour indices}
              \label{fig:tri}
    \end{figure}

Fig.~\ref{fig:tri} shows the evolution of the J-H, J-K and J-L
colour indices, which are strongly influenced by the presence of dust
around the system. Their maxima clearly indicate the epochs of
maximum obscuration.


\begin{figure}
      \resizebox{\hsize}{!}{\includegraphics[angle=-90]{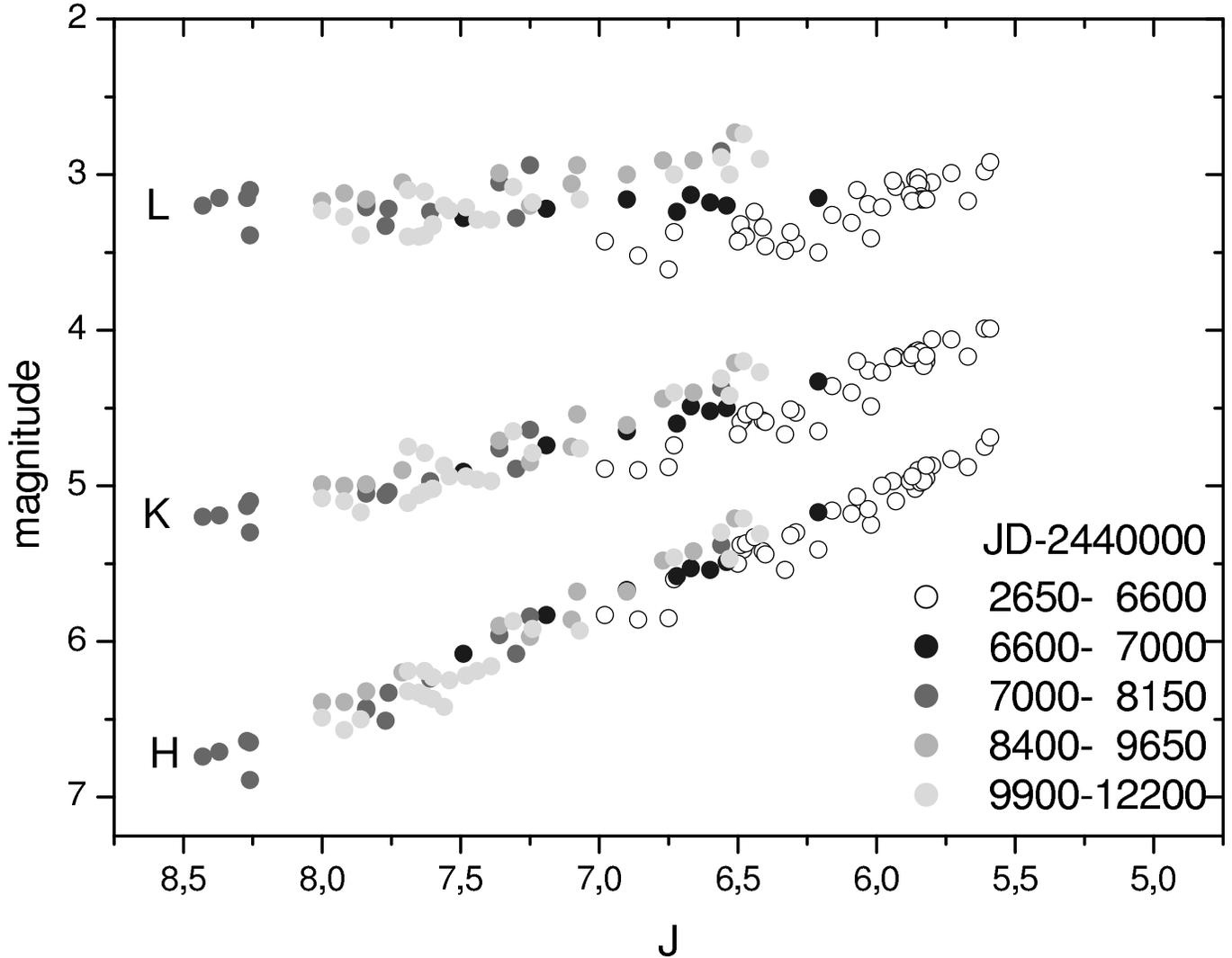}}
      \caption{Correlation of the H, K and L magnitudes with J for
different time periods. Empty symbols refer to dates before JD
2446600.  The transition epoch (JD 2446600-2447000) characterized
by the gradual change in the behaviour of the colour variations
and the following three obscuration events I, II and III are
distinguished by different grey tones.}
              \label{fig:cet}
    \end{figure}

More detailed information on the colour behaviour of RR Tel,
concerning differences between the three obscuration phases and
the epochs preceding them, can be derived from Fig.~\ref{fig:cet}.
The best statistical fit to the J, H, K, L correlations indicates
no difference among the correlations during the three obscuration
events in each band. The correlation gradients have the same sign
in all bands, being smaller when the wavelength with which J is
correlated is larger. This is to be expected in the presence of
absorption due to dust, where the absorption will decrease with
wavelength. The correlations suggest a change in colour behaviour
around JD2446600, the colours becoming redder in the later period.
The change, which was not sudden is clearest in the L band where
RR Tel is brightest. This could  result from grain fragmentation
and grain vaporization which will create an excess of small dust
particles in a low velocity (V$<$ 200 km/s) shock regime compared
to the unshocked one. However, we would need to have infrared
spectra for the same epochs to be sure that no other effect is
involved.

\begin{table*}
\caption{The features and time correlation of the IR and visual
light curves} \label{table:2}
\begin{tabular}
[c]{|l|l|l|l|l|l|l|l|l|l|}\hline\hline \multicolumn{6}{|c|}{IR
photometry} & \multicolumn{2}{|c|}{time} &
\multicolumn{2}{|c|}{visual photometry}\\\hline epoch &
\multicolumn{4}{c|}{$\Delta$m} & phase & JD-2440000 & date & epoch
&
m\\\cline{2-5}\cline{7-8}\cline{9-10}%
& J & H & K & L &  &  &  &  & \\\cline{1-6}\cline{7-8}\cline{10-10}%
&  &  &  &  &  & 4150 & Oct 1979 & a & 10,00\\\cline{1-6}\cline{7-8}%
\cline{10-10}%
&  &  &  &  &  & 5300 & Nov 1984 &  & 10,30\\\cline{1-6}\cline{7-8}%
\cline{10-10}%
&  &  &  &  & start & 6600 & Jun 1986 &  & 10,60\\\cline{6-8}\cline{9-10}%
I & 2,6 & 2,0 & 1,1 & 0,2 &  & 7000 & Aug 1987 &  & 10,75\\\cline{6-8}%
\cline{9-10}%
&  &  &  &  & minimum & 7700 & Jun 1989 &  & 10,75\\\cline{1-6}\cline{6-8}%
\cline{10-10}%
&  &  &  &  & start & 8400 & May 1991 & b & 10,75\\\cline{6-8}\cline{10-10}%
II & 2,2 & 1,5 & 0,8 & 0,0 & minimum & 8800 & Jun 1992 &  & 10,75\\\cline{1-6}%
\cline{7-8}\cline{9-10}%
&  &  &  &  &  & 10000 & Nov 1995 &  & 10,75\\\cline{1-6}\cline{7-8}%
\cline{9-10}%
&  &  &  &  & start & 10200 & May 1996 &  & 10,75\\\cline{6-8}\cline{10-10}%
III & 2,2 & 1,7 & 1,0 & 0,2 & minimum & 11100 & Oct 1998 & c &
11,05\\\cline{6-8}\cline{10-10}%
&  &  &  &  &  & 12440 & Jun 2002 &  & 11,50\\\hline
\end{tabular}
\end{table*}

\begin{figure}
      \resizebox{\hsize}{!}{\includegraphics[angle=-90]{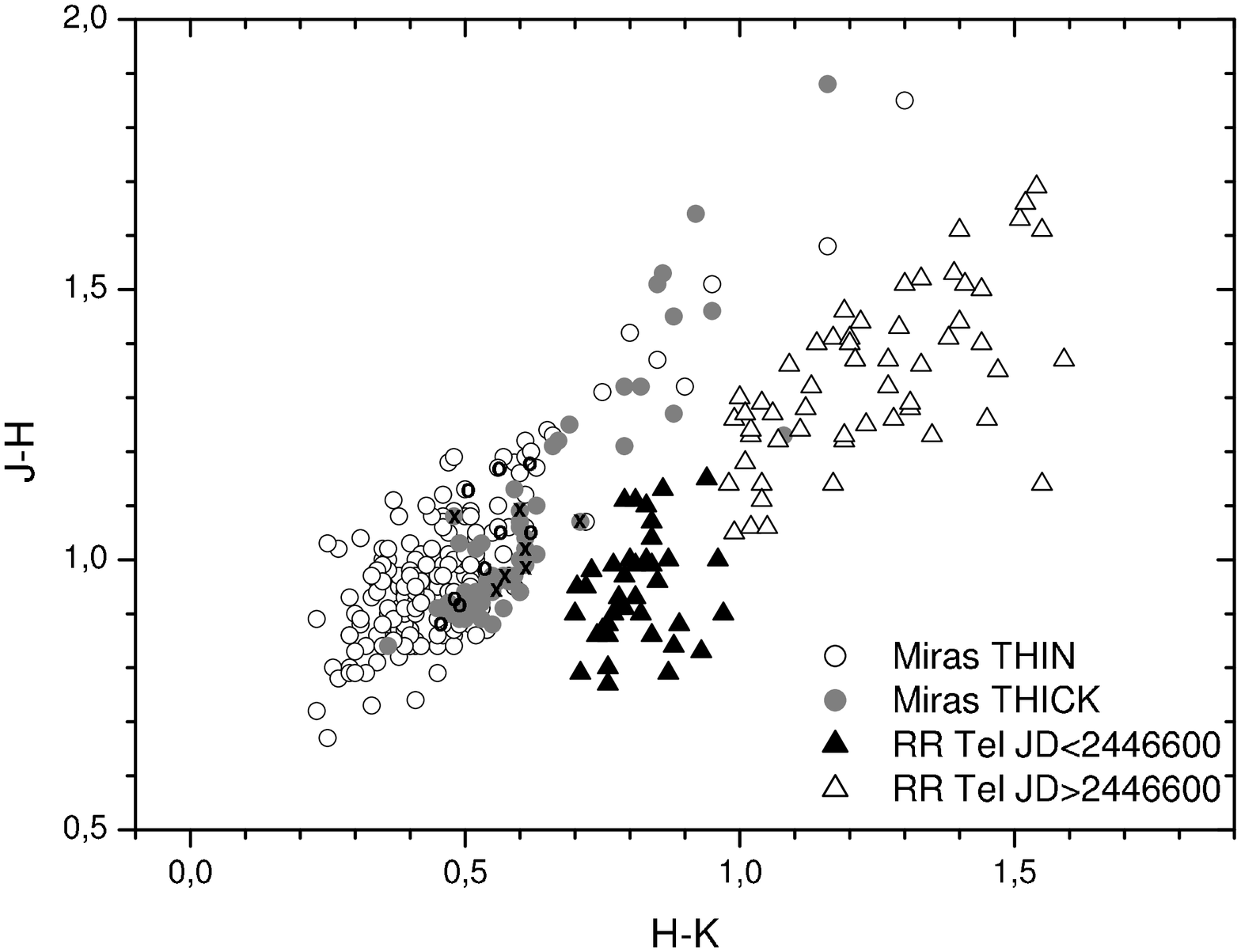}}
      \caption{Two colour  J-H,H-K diagram of RR Tel
  before (full triangles) and after (empty triangles) JD 2446600,
  as well as of normal Miras with thin dust shells (empty circles)
  from Whitelock et al. (\cite{WM00})
  and with thick dust shells (full circles) from Whitelock et al. (\cite{Wh94}) .
  The Miras with similar periods of pulsation to that of RR Tel are distinguished by crosses
  for thick dust shells and by bold circles for thin dust shells.
  Straight lines are least square fits for RR Tel before and after JD 2446600.}
              \label{fig:pet}
    \end{figure}

\begin{figure}
      \resizebox{\hsize}{!}{\includegraphics[angle=-90]{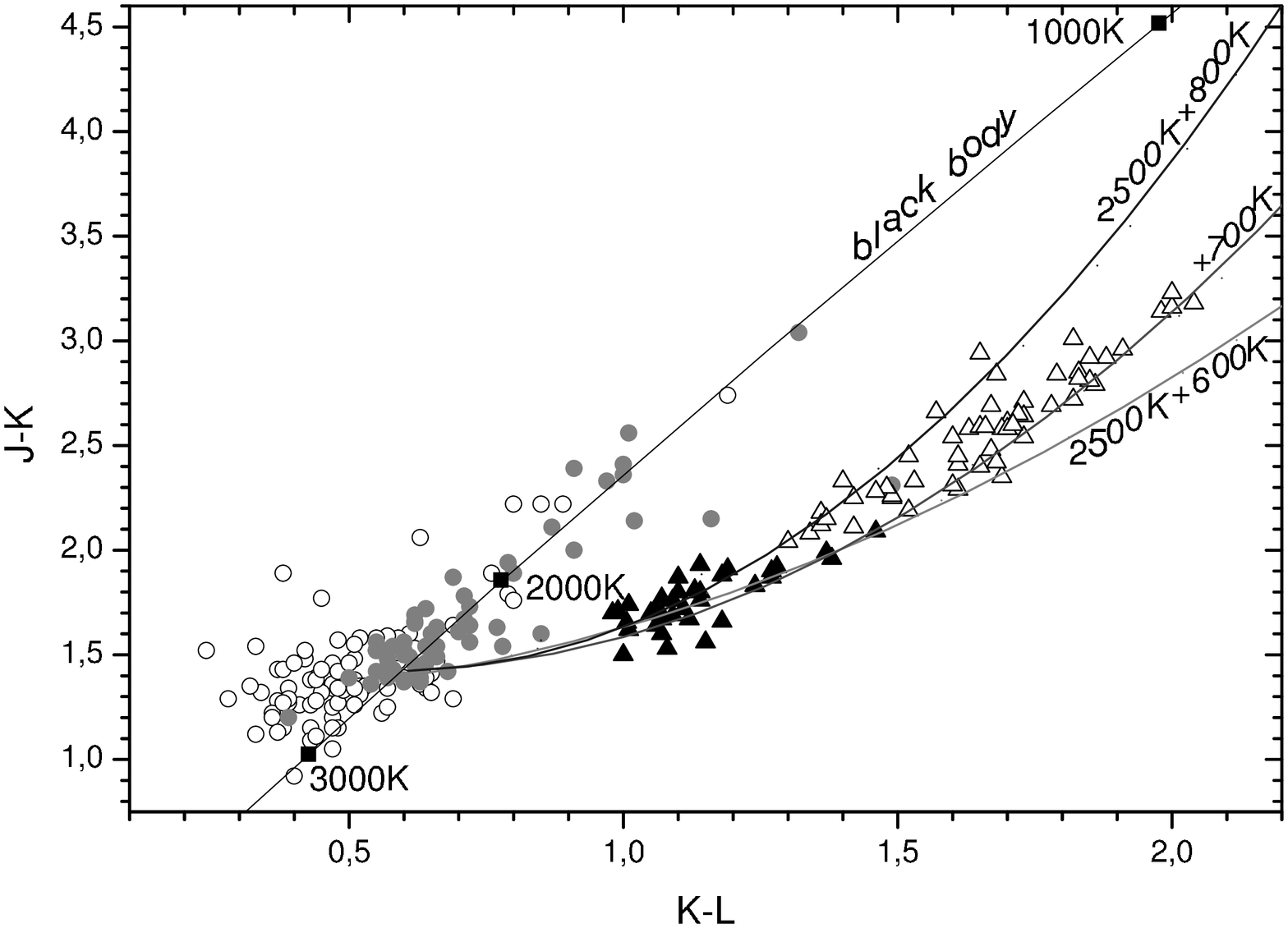}}
      \caption{Two colour  J-K,K-L diagram of RR Tel. In addition to
  the same symbols as in Fig.5,  the black body curve and
and curves representing a 2500 K Mira + 800 K, 700 K and 600 K
silicate dust shell of variable optical depth are plotted.}
              \label{fig:sest}
    \end{figure}

It is instructive to compare the colours of RR Tel with the mean
values of normal Miras. The RR Tel J-H,H-K and J-K,K-L two colour
diagrams as well as those for Miras with mainly thin dust shells
(Whitelock et al. \cite{WM00}) and for IRAS selected Miras with
relatively thick dust shells (Whitelock et al. \cite{Wh94}) are
shown in Fig. ~\ref{fig:pet} and Fig. ~\ref{fig:sest}. It is
clear that the observed RR Tel infrared colours are
significantly shifted to the right of the range shown by normal
Miras.

The colours of RR Tel can be explained by reddening which
dominates over a small amount of dust emission. This
interpretation was discussed by Whitelock \cite{Wh87}, who
presented a sample of symbiotic binaries in a two-colour diagram,
modeling their colours with a 2500K star plus an 800K dust shell
(represented as a locus on her Fig. 3). She discussed the mean
colours of six symbiotics in and out of obscuration events and
noted that they tended to move higher up the locus during the
events. This is consistent with the obscuration being caused by
increased dust absorption in the line of sight. The locus,
represented in our two-colour J-K,K-L diagram, was calculated on
the assumption that silicate dust surrounds the Mira whose radiation
it absorbs and re-emits at the given temperature. The further up
the locus, the thicker is the dust shell. RR Tel follows the same
type of trend as the other objects showing obscuration events.

The changes in the colours of RR Tel are consistent with their
being caused by changes in the optical depth of a silicate dust
shell (with grain properties from Le Bertre et al. (\cite{LB84})
table 2) which re-radiates predominantly around 700K (see Fig.
~\ref{fig:sest}).

This is of course very much an oversimplification, but it produces
a good first-order fit to the near-infrared colours. The
displacement of RR Tel, like that of other symbiotic Miras, with
respect to the single Miras, is according to the present model, a
consequence of the higher temperature (600-800 K) of the dust
shell around the cool component (Danchi et al. \cite{Da94}),
compared with temperatures 100-300 K of non-symbiotic Miras
(Whitelock \cite{Wh87}).

We wonder if the extra heating is provided  by the hot component
of the binary system as long as its temperature is not too high to
evaporate the dust grains. Actually, the theoretical models of
Danchi et al. (\cite{Da94}) indicate relatively high dust
temperatures ($>$ 500 K) also in  dust shells around isolated
Mirae.

In Table 3 we compare the extinction in obscuration phase I
 with that expected from
interstellar extinction according to the Rieke and Lebofsky law
(Rieke and Lebofsky \cite{RL85}) and the van der Hulst curve 15
(vdH15). We also show for comparison the extinction experienced by
the symbiotic Mira R Aqr (Whitelock et al. \cite{Wh83}) during its
1976 obscuration event. All values have been normalized to $\Delta
J=1$.

\begin{figure}
      \resizebox{\hsize}{!}{\includegraphics[angle=-90]{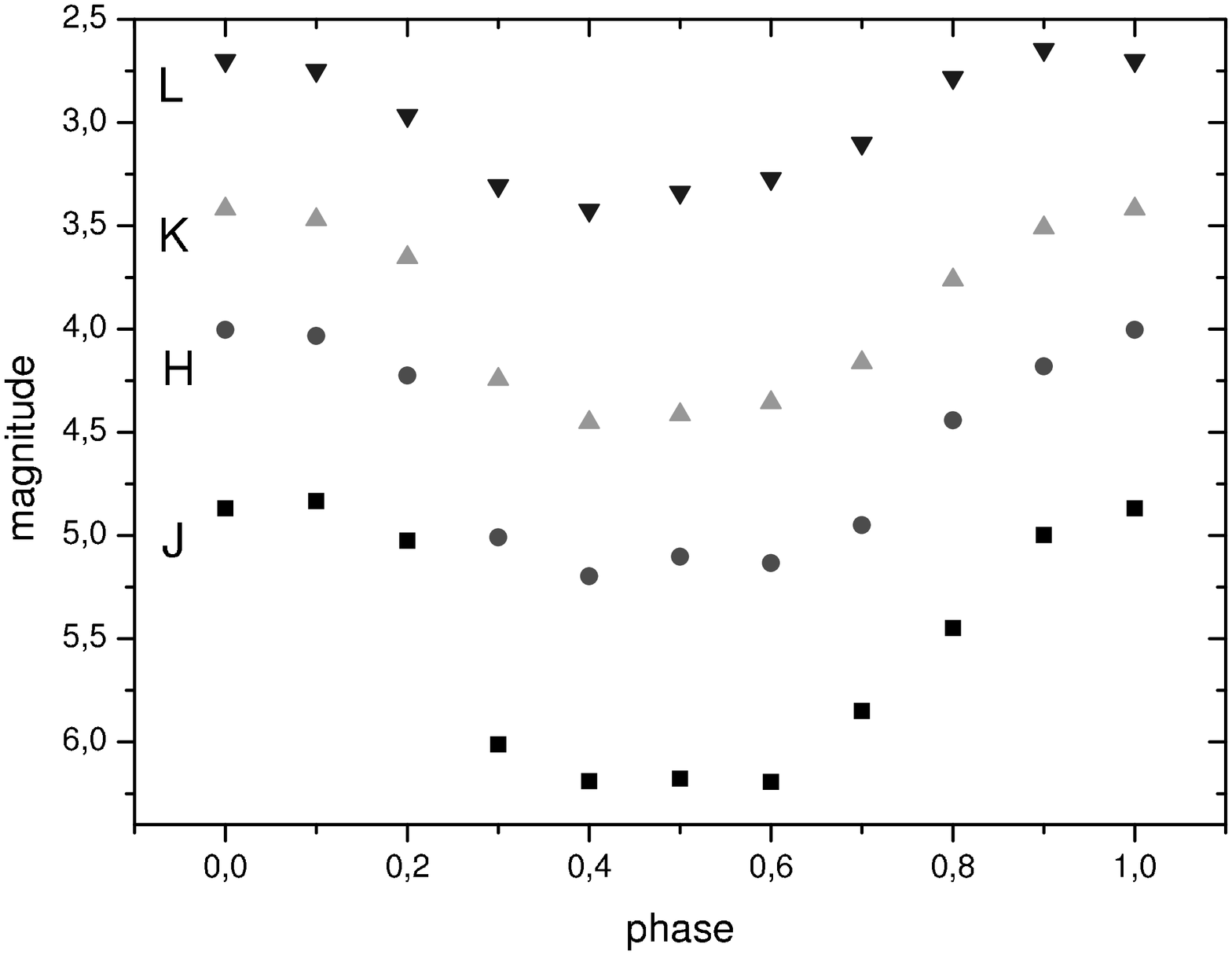}}
      \caption{Mira pulsations of RR Tel at different wavelengths}
              \label{fig:sedam}
    \end{figure}

\begin{figure}
      \resizebox{\hsize}{!}{\includegraphics[angle=-90]{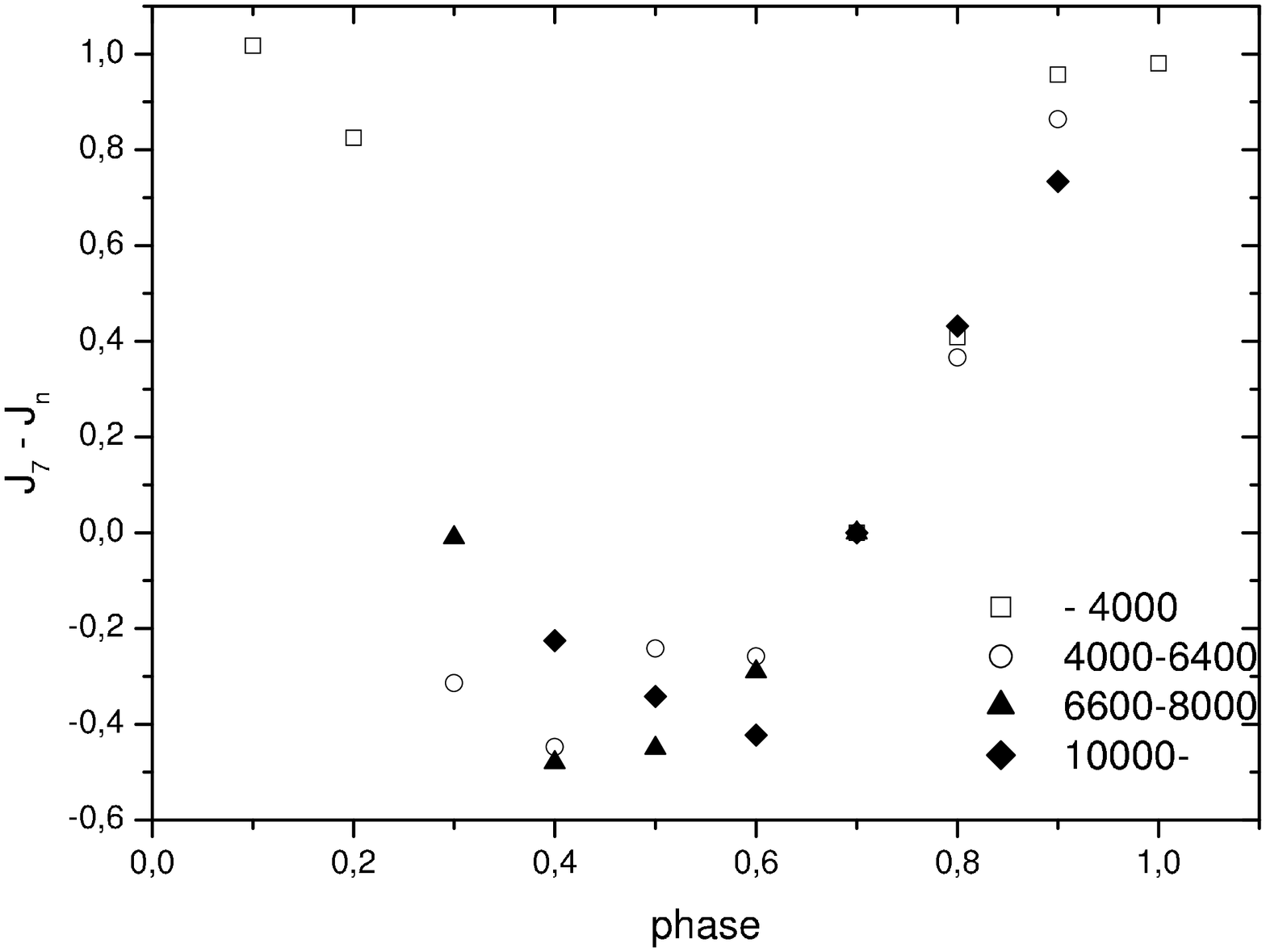}}
      \caption{The Mira pulsations  in J at different epochs before
  the start of obscuration event I at JD 2446600
  (empty symbols)  and after it (full symbols): till JD
  2444000 (squares), JD 2444000-JD 2446400 (circles), JD 2446600-JD 2448000
  (triangles), after JD 2450000 (diamonds). The magnitudes at each
  phase J$_7$-J$_n$ are given relative to that at phase 0.7.}
              \label{fig:osam}
    \end{figure}

A comparison of the figures in Table 3 suggests that the obscuration at JHK
is plausibly caused by extinction by particles comparable to those found in
the interstellar medium. However, the fact that the extinction at L is
rather less than predicted by normal reddening laws and is unlike the
case of \object{R Aqr}, may suggest somewhat different grain sizes.

\begin{table}
\caption{Wavelength dependence of extinction during obscuration
phases} \label{table:2}
\centering
\begin{tabular}{ccccc}
\hline\hline $\Delta J$&$\Delta H$&$\Delta K$&$\Delta L$& source\\
\hline
 1.0 & 0.77 &
0.42 & 0.08 & RR Tel phase I (this work)\\ 1.0 & 0.63 & 0.36 & 0.18 & vdH 15 (Rieke and Lebofsky \cite{RL85})\\
1.0 & 0.80 & 0.51 & 0.26 & Rieke and Lebofsky \cite{RL85} \\ 1.0 &
0.8 & 0.6 & 0.3 & R Aqr 1976 (Whitelock et
al. \cite{Wh83})\\
\hline
\end{tabular}
\end{table}

Comparison of the amplitudes of variation (Fig.~\ref{fig:sedam})
with the mean colours shows no clear difference between the Mira
component of RR Tel and single Miras. Particularly RR Tel fits
well into the correlation between the pulsation amplitude in the K
band and K-L' of Le Bertre (\cite{Be93}), but shows a somewhat
larger K pulsation amplitude relative to all the oxygen-rich
late-type stars of the sample in the period vs. K pulsation
amplitude correlation. Olivier et al. (\cite{Ol01}) also studied
the amplitudes of variation of dust enshrouded AGB stars. The
amplitude of the Mira component of RR Tel is larger than that
expected for stars of about the same pulsation period according to
their period-amplitude correlation, except for one C-rich object
which shows about the same deviation in the L band. The Mira
pulsation amplitudes in the two correlations between amplitudes in
different bands appear to be fairly normal according to the values
of Smith (\cite{Sm03}).

 The contribution of other non-Mira sources of radiation to the infrared
magnitudes and especially to J, is not easy to see. One way to
study it, would in principle be to compare the apparent amplitude
of the Mira pulsations during obscuration events to that outside
obscuration events. The proportion of radiation coming from the
Mira would be less at Mira minimum during an obscuration event
than outside it, so decreasing the apparent Mira amplitude.
Fig.~\ref{fig:osam} shows the Mira J light curve, which appears to
be similar at different times. However, the method is in our case
not reliable enough because the Mira light-curve interpolation
procedure does not work very well during obscuration events.
Nevertheless, we doubt that the hot component is responsible for
the systematic effects in the near IR. One reason is the lack of
evidence of a strong continuum in the optical and UV (Penston et
al. \cite{Pe83}).

The nebular component will contribute mainly through the emission
lines and through the bremsstrahlung which  dominates in the radio
range.

The infrared spectrum of RR Tel (Feast et al. \cite{FW83}) shows
strong water absorption without obvious emission lines and so
suggests that at a Mira phase of 0.8 (Belczynski et al.
\cite{Be00}) the radiation from the cool component is the major
contributor in J.

Other evidence for a negligible contribution from anything
other than the Mira to the J flux is the large amplitude of the
dust obscuration events seen in Fig.~\ref{fig:dva}. The largest J
amplitude is 2.6 magnitudes (factor of 11), with other events of
smaller amplitude. On the other hand, the strongest absorption
feature due to water vapour presumably due to the Mira, seen in
the rather low resolution infrared spectrum of Feast et al.
(\cite{FW83}) at 1.4 $\mu$m, has an apparent depth of about 60
percent of the continuum. That spectrum of 1980 Aug 20 was taken
when dust obscuration was near minimum. The disappearance of all
the flux below the absorption feature, even if one supposes it
only due to the nebular component, would limit the effect of dust
obscuration to 0.55 magnitudes, assuming that obscuration does not
affect this nebular component.

\subsection{Optical Fe II emission lines}

The fluxes of the permitted Fe II and forbidden [Fe II] lines, in
the spectra taken in 1996 and 2000, were measured and compared.
The spectra taken in 2000 were corrected for reddening using the
reddening law of Howarth (\cite{Ho83}) with R=3.1 and E(B-V)=0.08.
Crawford corrected the 1996 spectra for interstellar extinction
using the coefficients listed in Cardelli et al. (\cite{Ca89}) and
E(B-V)=0.08.

\begin{figure}
      \resizebox{\hsize}{!}{\includegraphics[angle=-90]{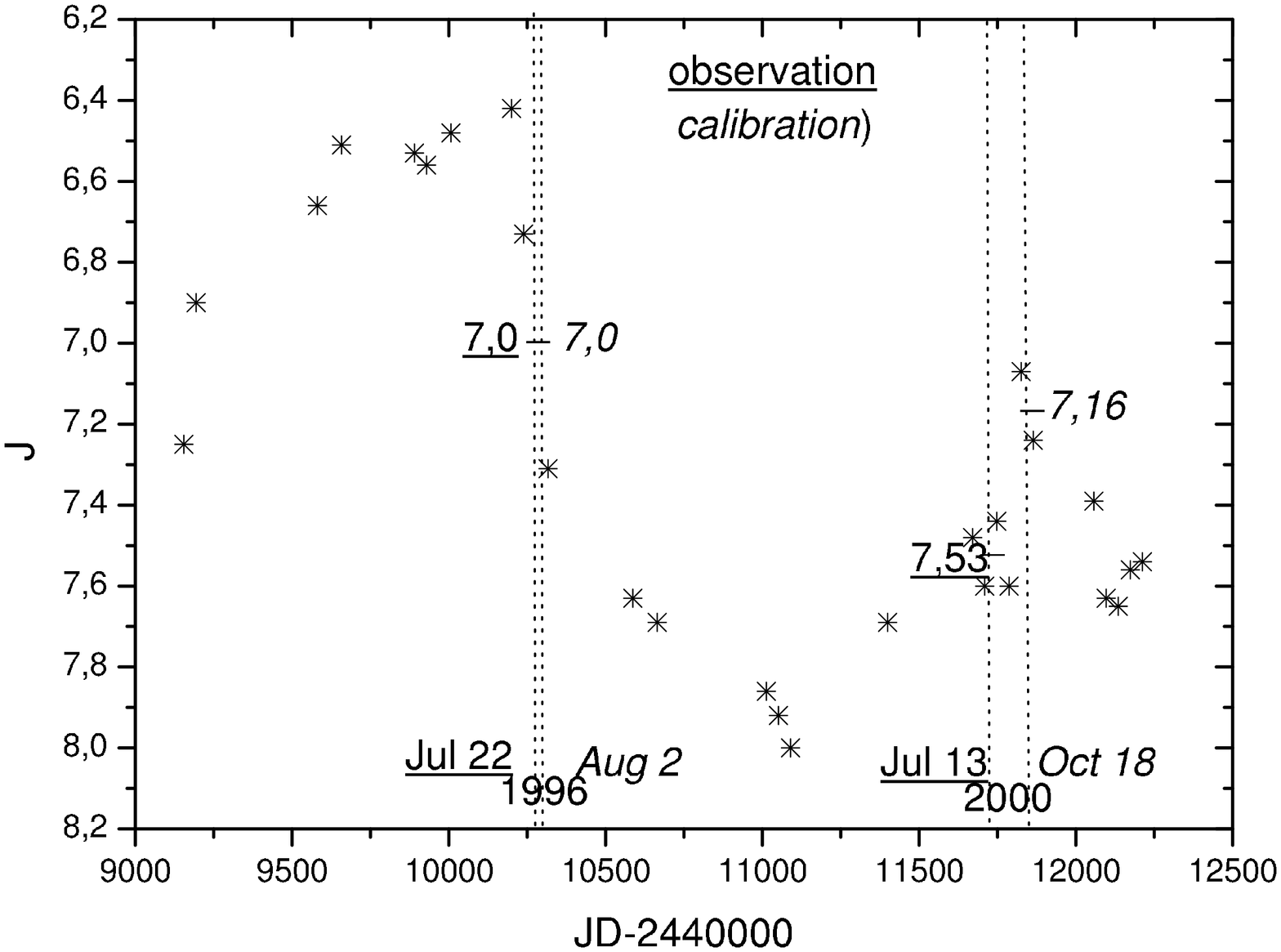}}
      \caption{J light curve corrected for Mira pulsations during the
  dust obscuration episode when the two sets of optical spectra, 1996 and 2000, were taken}
              \label{fig:devet}
    \end{figure}

\begin{figure}
      \resizebox{\hsize}{!}{\includegraphics[angle=-90]{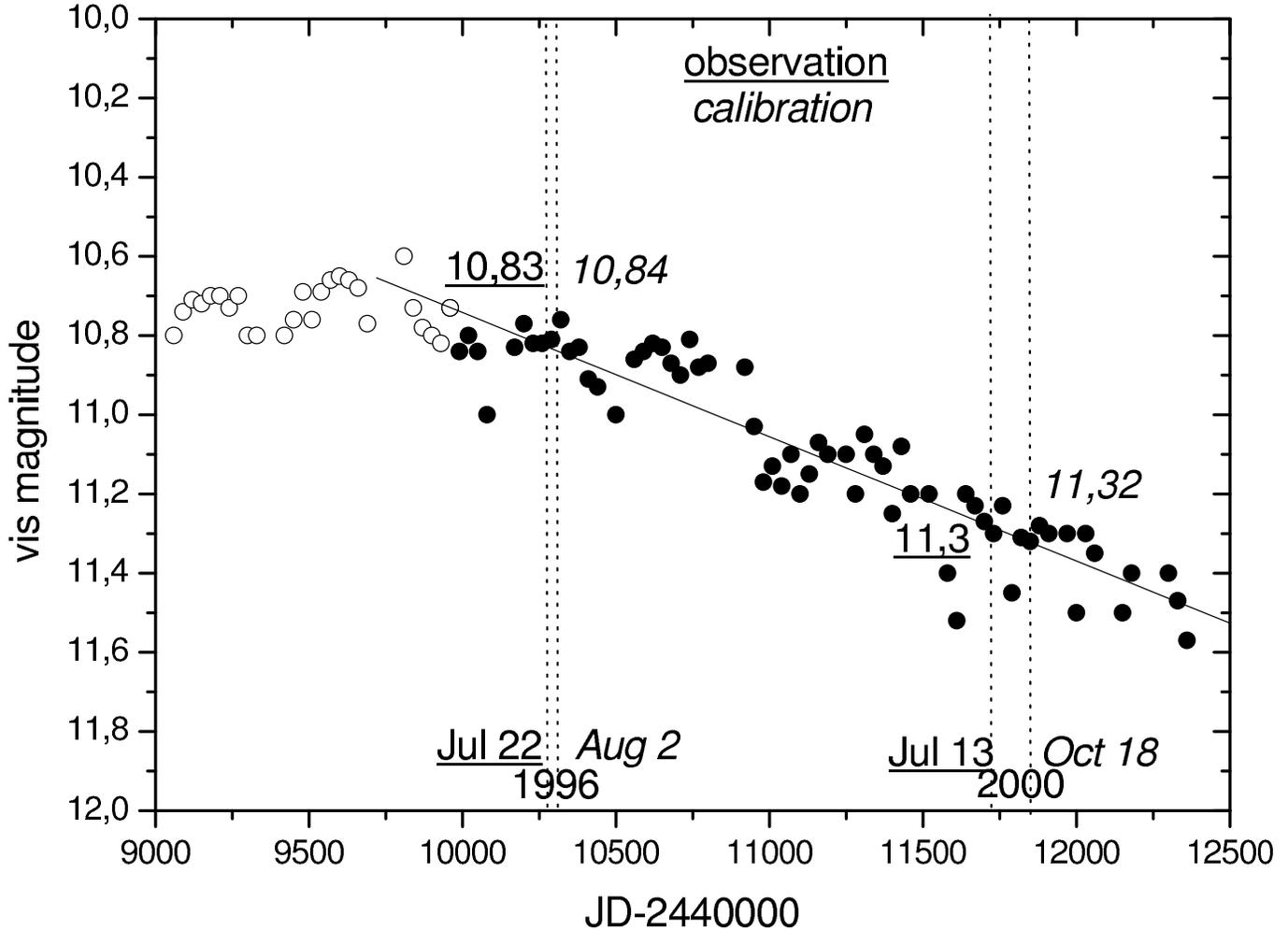}}
      \caption{Visual light curve of RR Tel during the
  dust obscuration episode when the two optical spectra, 1996 and 2000, were taken}
              \label{fig:deset}
    \end{figure}

The dates of the spectra are marked in the corresponding segment
of the J and visual light curves which are plotted in Figs.
~\ref{fig:devet} and ~\ref{fig:deset} respectively. A dust
obscuration event was just starting when the 1996 spectra were
taken and underway when the 2000 spectra were obtained. Correcting
for the Mira pulsations, a fading of only 0.16 magnitudes in J is
obtained between the times of the calibration spectra, because of
a temporary J brightening when the calibration of the later
spectrum was obtained (Fig.~\ref{fig:devet}). The observed visual
fading between the same two dates was 0.48 mag
(Fig.~\ref{fig:deset}).

It might be interesting to compare the fading in V due to many
contributions, depending on changing physical conditions as well as
on the obscuration event, with the Fe II line and infrared fadings.

Other lines from different levels (Kotnik-Karuza et al.
\cite{KK04}, Kotnik-Karuza et al. \cite{KF04}) behave differently
because they are sensitive in different ways to the physical
conditions of the emitting gas.

According to an interstellar extinction law (Rieke and Lebofsky
\cite{RL85}), the fading of 0.16 in J would correspond to a fading
of 0.57 mag in V.

\begin{figure}
      \resizebox{\hsize}{!}{\includegraphics[angle=-90]{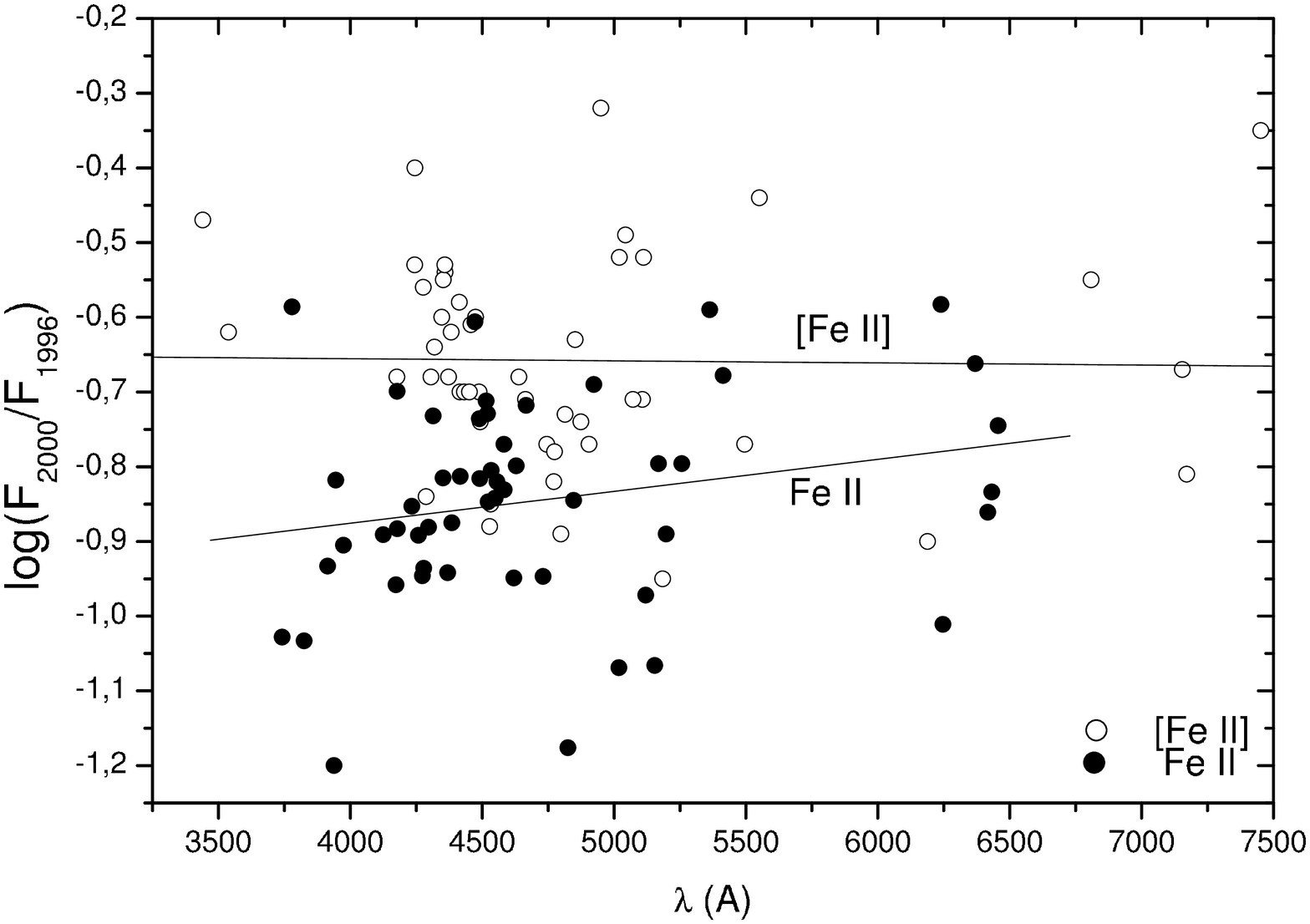}}
      \caption{Fading of Fe II and [Fe II] log line fluxes from 1996
  to 2000 as a function of wavelength}
              \label{fig:enast}
    \end{figure}

The latter corresponds to a flux decrease given by a log flux
ratio equal to -0.23 in the V band if the radiation in J and V
came from the same region and was absorbed by the same medium. Log
flux ratios between the two dates for the Fe II and [Fe II]
emission line fluxes in the two spectra, plotted against
wavelength, are shown in Fig.~\ref{fig:enast}. Lines with
wavelengths below 3490 \AA~ were not taken into account in our
iron  line study, as the Crawford calibration in that region is
highly uncertain. The measured line of multiplet 73 at a large
wavelength of 7711 \AA~ and a few other weak extremely discordant
lines were also eliminated. Only the difference between the
permitted and forbidden lines appears to be significant, showing a
larger flux decrease for the permitted than for the forbidden
lines. No significant correlation has been found between other
line properties and log ratio. The best correlation between the
log flux ratio of the forbidden lines and log $(gf \lambda)$ still
has a probability of 0.25 and would moreover be difficult to
understand for optically thin forbidden lines. The mean log flux
ratios are $-0.85 \pm 0.14$ for Fe II and $-0.66 \pm 0.14$ for [Fe
II].

Note that the log flux change of the iron lines between the two
spectra is much larger than the change in the V band over the same
period. It therefore seems possible that these lines are formed
closer to the Mira than the higher temperature region responsible
for most of the emission at V.

 Let us note that the narrowness of the Fe II and [Fe II] emission lines,
found by Kotnik-Karuza et al. (\cite{KK02}), indicates line
formation in a low velocity wind of the cool component.

Looking for the simplest possible interpretation of what we see
and especially Fig.~\ref{fig:enast}, two types of model might be
considered. One involves spherical symmetry of the extra dust
absorption in the Mira wind, while the other assumes the presence
of one or more clouds. In both cases the apparent radius of the
line emitting region found by the Self-Absorption Curve (SAC)
method, might be expected to decrease during a dust obscuration
episode (Kotnik-Karuza et al. \cite{KK02}, Kotnik-Karuza et al.
\cite{KK03}). In both cases the presumably less absorbed forbidden
line region will be larger.

\par The log optical absorption of 0.23, corresponding to the J
absorption, is less than that of the Fe II lines. One might think
that this disagreement is due to the difficulty of finding the
exact J magnitudes when the spectra were taken, so the spherically
symmetric model, with fewer grains above the forbidden line region
than the permitted line one, is not then necessarily contradicted.

We must, nevertheless, emphasize that the low velocity of the Mira
wind should lead to the line formation regions being occulted by
dust significantly later than the Mira itself, so direct
comparison of the J and line flux fadings may be misleading.

\par The lack of a clear wavelength dependence of the
extra absorption of the optical lines, unlike the behaviour shown
in the near infra-red, is, however, a problem and might suggest
rather the presence of one or more optically thick clouds
occulting much of the Fe II line emitting regions.

\subsection{Distance estimates}

\bigskip

Studies of the optical emission spectrum of RR Tel suggest
considerably more dust toward the cool star than on the line of
sight to the high excitation regions (Kotnik-Karuza et al.
\cite{KF04}). Thus, in determining the distance to the star, one
should take into account the fact that the extinction in the
direction of the cool component includes circumstellar as well as
interstellar extinction. The interstellar extinction towards RR
Tel is $E_(B-V)=0.10-0.11$ (Penston et al. \cite{Pe83}; Young et
al. \cite{Yo05}) which would give $A_K \sim 0.03$ mag.

For the distance estimate we use the mean values of the K
magnitude and of the J-K colour index equal to 4.16 and 1.69,
respectively, obtained from observations at the epoch preceding
the first obscuration event (I), when the circumstellar absorption
was at its minimum.

From the value of 387 days for the period of Mira pulsations and
also using the correlation between the Mira period and absolute
magnitude (Feast et al. \cite{FG89}, assuming a distance modulus
for the \object{LMC} of 18.5 mag), we obtain an absolute magnitude
in K, M$_K= -8.0$. The unabsorbed J-K was obtained by applying the
period-colour relation for Mira variables in the solar
neighbourhood from Whitelock et al. (\cite{WM00}), $
(J-K)_0=-0.39+0.71 \log P $. This leads to $(J-K)_0=1.45$ and
$E_{J-K}=0.24$. Using the reddening law specified by Glass
(\cite{Gl99}) we find A$_K=0.13$ - indicating that most of the
obscuration is circumstellar rather than interstellar.

Thus the distance to RR Tel is 2.5kpc, which is very close to the
value of 2.6 kpc found e.g. by Whitelock (\cite{Wh88}). This
procedure assumes that the circumstellar extinction has the same
reddening characteristics as does interstellar extinction. In view
of the discussion of the extinction during the obscuration phases
(section 3.2) and the fact that the correction is small, the
assumption is unlikely to cause serious errors in the distance,
which should be good to about 0.3 kpc.

\section{Conclusions}

   Our study, of both broad band fluxes and the fluxes of emission
lines of once ionized iron, lead to suggestive results, whose
interpretation is, however, still uncertain. The infrared flux
variations are understandable in terms of wavelength dependent
absorption by dust around the cool component. It is not clear what
the reason is for the change in behaviour of fluxes between JD
2446600 and JD 2447000. The effect with respect to J is larger
for longer wavelength bands, suggesting that the continuum
absorption of dust is responsible, rather than changes of emission
line fluxes.

The two-colour diagrams show unusual characteristics for RR Tel
compared with the colours of single Miras. The track during the
fading events is consistent with obscuration caused by an
increasingly thick dust shell. The characteristic temperature of
the dust shell, 700K, is sufficiently hot to produce an excess in
the L band which results in the observed colours. A qualitatively
similar explanation was offered for the colours of symbiotic Miras
by Whitelock (\cite{Wh87}).

The relatively high temperature found for the dust shell could be
a consequence of heating by the binary companion. A consistent
model is needed to check this hypothesis.

The optical magnitudes are not simply correlated with the infrared
ones, but later fading might be connected with dust obscuration.
It should be noted that the optical behaviour is governed by the
behaviour of the different emission line fluxes.

The apparently wavelength independent variation of the fluxes of
the lines of ionized iron during dust obscuration is most simply
understood as being due to absorption by separate optically thick
clouds.

If the narrow ionized iron lines (Kotnik-Karuza et al.
\cite{KK03}) are formed in the cool star's wind, the lines could
be formed farther out from the cool component than the dust
producing the infrared broad band absorption. If fewer clouds
absorb, the then less absorbed forbidden lines appear to be, as
expected, formed farther out than the permitted lines. However, we
do not have enough observations at different epochs; ionized iron
and perhaps other emission line fluxes should rise at the end of
obscuration events, if the present interpretation is correct. We
should note that the presence of separate clouds is not
unexpected, as it is rather similar to what is thought to occur
during fadings of \object{R Corona Borealis} stars (Feast
\cite{Fe86}).

\begin{acknowledgements}
      We are grateful to Michael Feast for starting the IR monitoring
programme at SAAO. Our thanks are also due to the late Janet
Mattei, Rebecca Pollock and to Albert Jones for providing us with
the optical photometry of RR\,Tel.
\end{acknowledgements}

{}

\end{document}